\documentclass{article}
\usepackage{amssymb}

\usepackage{graphicx}
\usepackage{amsmath}

\newtheorem{theorem}{Theorem}

\newtheorem{corollary}[theorem]{Corollary}

\newtheorem{definition}[theorem]{Definition}

\newtheorem{exercise}[theorem]{Exercise}

\newtheorem{notation}[theorem]{Notation}

\newtheorem{remark}[theorem]{Remark}

\newenvironment{proof}[1][Proof]{\textbf{#1.} }{\ \rule{0.5em}{0.5em}}
\input{tcilatex}

\begin{document}

\title{A new Quantum Mechanics in Phase Space}
\author{\textbf{Antonio Cassa} \\
antonio.leonardo.cassa@gmail.com}
\maketitle

\begin{abstract}
For each bounded operator $A$ on the Hilbert space $L^{2}(\mathbb{R}^{m})$
we define a function $\left\langle A\right\rangle :\mathbb{R}%
_{qp}^{2m}\rightarrow \mathbb{C}$ taking on $(q,p)$ the expected values of $%
A $ on a suitable state $\vartheta _{qp}$. For $A\geqslant B$ we have $%
\left\langle A\right\rangle \geqslant \left\langle B\right\rangle $. Dually
for each couple $\varphi .\psi $ we define a function $S_{\varphi \psi }$ on
the $\left\langle A\right\rangle $ in such a way to have $S_{\varphi \psi }($
$\left\langle A\right\rangle )=\left\langle \varphi ,A\psi \right\rangle $.
\end{abstract}

\section{Introduction}

\bigskip

\bigskip This paper developes a new version of the Quantum Mechanics on
Phase Space (cfr: [\textbf{CZ}], [\textbf{dG}], [\textbf{Gr}], [\textbf{K}],
[\textbf{M}] or [\textbf{P}]) associating a (restricted) expected value
function $\left\langle A\right\rangle $ to each bounded self-adjoint
operator $A$ and a ''distribution'' $S_{\varphi \psi }$ on $\mathbb{R}^{2m}$
to each couple of vector states in such a way to have: $S_{\varphi \psi
}(\left\langle A\right\rangle )=\left\langle \varphi ,A\psi \right\rangle $ (%
$S_{\varphi \psi }$ is a linear function on the space $\mathcal{M}$ of all
the functions $\left\langle A\right\rangle $)

The new symbol $\left\langle A\right\rangle $ introduced here has the
advantage to be positive when $A$ is positive.

The terms $\left\langle A\right\rangle $ and $S_{\varphi \psi }$ are
connected with the Wigner and\ Husimi \ transforms (cfr. Remark $[17]$). We
give explicit formulas for $\left\langle A\right\rangle (q,p)$, $S_{\varphi
\psi }(\left\langle A\right\rangle )$ and for the product in $\mathcal{M}$.

\bigskip

\section{Symbols}

\bigskip As usual $\mathcal{E}(\mathbb{R}^{m})$ will denote the space of all
differentiable functions on $\mathbb{R}^{m}$. On the space of square
integrable functions $L^{2}(\mathbb{R}^{m})$ we will use the convolution
operation $\ast :L^{2}(\mathbb{R}^{m})\times L^{2}(\mathbb{R}%
^{m})\rightarrow L^{2}(\mathbb{R}^{m})$ given by: $\left( f\ast g\right) (x)=%
\frac{1}{\left( 2\pi \right) ^{m/2}}\cdot \int f(a)\cdot g(x-a)\cdot da$
(note the coefficient) and the Fourier transform $\mathcal{F}:L^{2}(\mathbb{R%
}^{m})\rightarrow L^{2}(\mathbb{R}^{m})$ given by $\mathcal{F}(f)(\xi )=%
\frac{1}{\left( 2\pi \right) ^{m/2}}\cdot \int_{\mathbb{R}^{m}}f(x)\cdot
e^{-i\cdot x\cdot \xi }\cdot dx$. With these positions we have exactly $%
\mathcal{F}\left( f\ast g\right) =\mathcal{F}\left( f)\cdot \mathcal{F(}%
g\right) $. On the space $L^{2}(\mathbb{R}_{x}^{m}\times \mathbb{R}_{y}^{m})$
we will consider also the partial Fourier transform $\mathcal{F}_{I}:L^{2}(%
\mathbb{R}^{2m})\rightarrow L^{2}(\mathbb{R}^{2m})$ given by $\mathcal{F}%
_{I}(F)(u,y)=\frac{1}{\left( 2\pi \right) ^{m/2}}\cdot \int_{\mathbb{R}%
^{m}}F(a,y)\cdot e^{-i\cdot a\cdot u}\cdot da$ , analogously we define $%
\mathcal{F}_{II}$. In this context we will meet the changement of variables $%
\tau :\mathbb{R}^{2m}\rightarrow \mathbb{R}^{2m}$ given by $\tau (x,y)=(y+%
\frac{x}{2},y-\frac{x}{2})$, its inverse $\tau ^{-1}(u,v)=(u-v,\frac{u+v}{2}%
) $ and the exchange map $\Xi :\mathbb{R}^{2m}\rightarrow \mathbb{R}^{2m}$
given by\ $\Xi (x,y)=(y,x)$. We will also reserve the symbol $\omega $ to
the function $\omega :\mathbb{R}_{x}^{m}\times \mathbb{R}_{y}^{m}\rightarrow 
\mathbb{R}^{+}$ given by $\omega (x,y)=\frac{1}{\pi ^{m}}\cdot e^{-\left(
\left\| x\right\| ^{2}+\left\| y\right\| ^{2}\right) }$.

We will denote by $SK:L(\mathcal{H})\rightarrow \mathcal{S}^{^{\prime }}(%
\mathbb{R}^{2m})$ the (injective) Schwartz kernel map characterized by $%
\left\langle \varphi ,A\psi \right\rangle =SK(A)\left( \overline{\varphi }%
\otimes \psi \right) $ whenever $\varphi ,\psi \in \mathcal{S}(\mathbb{R}%
^{m})$ (cfr. \textbf{[T]}). When $\psi $ is a unitary vector, we will denote
by $E_{\psi }$ its associated projector on the line $\mathbb{C\cdot \psi }$.

\ 

\section{A functional representation for bounded operators}

\begin{definition}
\bigskip Taken $(q,p)$ in $\mathbb{R}^{2m}$ the \textbf{fundamental state
centered in }$(q,p)$ is the unitary vector $\vartheta _{qp}:\mathbb{R}%
^{m}\rightarrow \mathbb{C}$ given by: 
\begin{equation*}
\vartheta _{qp}(x)=\frac{1}{\pi ^{m/4}}e^{ip\cdot (x-\frac{q}{2})}\cdot e^{-%
\frac{1}{2}\left\| x-q\right\| ^{2}}
\end{equation*}
\end{definition}

\begin{remark}
Note that $\mathcal{F}\vartheta _{qp}=\vartheta _{p,-q}$. We will write: $%
\vartheta $ for $\vartheta _{oo}$.
\end{remark}

\begin{remark}
If we introduce the map $\mathcal{U}_{\cdot }:\mathbb{C}^{m}\rightarrow Unit(%
\mathcal{H})$ defined by $\mathcal{U}_{q+ip}(\psi )(x)=e^{ip\cdot (x-\frac{q%
}{2})}\cdot \psi (x-q)$ we have $\vartheta _{qp}=\mathcal{U}_{q+ip}\vartheta 
$. Note that $\mathcal{U}_{q+ip}(\psi )\sim \psi $ only for $(q,p)=(0,0)$.
\end{remark}

\begin{definition}
The \textbf{expected value on the fundamental states map} is the map $%
\left\langle \cdot \right\rangle :L_{sa}(\mathcal{H})\rightarrow \mathcal{E}(%
\mathbb{R}^{2m})$ given, for each self-adjoint bounded operator $A$ on $%
\mathcal{H}$, by $\left\langle A\right\rangle (q,p)=\left\langle \vartheta
_{qp},A\vartheta _{qp}\right\rangle =\left\langle A\right\rangle _{\vartheta
_{qp}}$

\begin{remark}
Whenever $A\geqslant 0$ we have $\left\langle A\right\rangle \geqslant 0$
and if $A\geqslant B$ we have $\left\langle A\right\rangle \geqslant
\left\langle B\right\rangle $; if $\left\{ E_{(-\infty ,r]}^{A}\right\}
_{r\in \mathbb{R}}$ is the spectral family of the bounded operator $A$ then
the function $\left\langle E_{(-\infty ,r]}^{A}\right\rangle (q,p)$ is, for
each $(q,p)$, a monotone non-decreasing function in $r$, right continuous,
with $\inf =0$ and $\sup =1$. The map $\left\langle \cdot \right\rangle $
extends to $\left\langle \cdot \right\rangle :L(\mathcal{H})\rightarrow 
\mathcal{E}_{\mathbb{C}}(\mathbb{R}^{2m})$ as a $\mathbf{C-}$linear map.
\end{remark}
\end{definition}

\begin{exercise}
\begin{itemize}
\item  $\left\langle g(Q_{k})\right\rangle (q,p)=\sqrt{2\pi }\cdot (g\ast
\vartheta ^{2})(q_{k})$ for each $g$ bounded

\item  $\left\langle f(P_{k})\right\rangle (q,p)=\sqrt{2\pi }\cdot (f\ast
\vartheta ^{2})(p_{k})$ for each $f$ bounded

\item  Note that: $\left\langle Q^{2}-\frac{1}{2}I\right\rangle (q,p)=\frac{1%
}{\sqrt{\pi }}q^{2}\geqslant 0$ but $Q^{2}-\frac{1}{2}I\ngeqslant 0$.

\item  $\left\langle E_{\psi }\right\rangle (q,p)=\left| \left\langle \psi
,\theta _{qp}\right\rangle \right| ^{2}$ for each unitary $\psi $

\item  Let $\left\{ E_{\psi _{0}},...,E_{\psi _{m},}...\right\} $ be a
sequence of pairwise orthogonal projectors in $\mathcal{H}$ such that $%
I=\sum_{k\geq 0}E_{\psi _{k}}$\ ; let $\lambda _{0},...,\lambda _{m},...$ be
a sequence of real numbers giving a bounded operator $A=\sum_{k\geq
0}\lambda _{k}\cdot E_{\psi _{k}}$ .We have: $\left\langle A\right\rangle
(q,p)=\sum_{k\geq 0}\lambda _{k}\cdot \left| \left\langle \psi _{k},\theta
_{qp}\right\rangle \right| ^{2}$
\end{itemize}
\end{exercise}

\begin{notation}
Denoted by $\mathcal{M}$ the image of the linear map $\left\langle \cdot
\right\rangle :L(\mathcal{H})\rightarrow \mathcal{E}_{\mathbb{C}}(\mathbb{R}%
^{2m})$, we will consider the following isomorphisms (as restricted maps): $%
SK:L(\mathcal{H})\rightarrow SK(L(\mathcal{H}))\subset \mathcal{S}^{\prime }(%
\mathbb{R}^{2m})$, $\tau ^{\ast }:SK(L(\mathcal{H}))\rightarrow (\tau ^{\ast
}\circ SK)(L(\mathcal{H}))\subset \mathcal{S}^{\prime }(\mathbb{R}^{2m})$, $%
\mathcal{F}_{I}:(\tau ^{\ast }\circ SK)(L(\mathcal{H}))\rightarrow (\mathcal{%
F}_{I}\circ \tau ^{\ast }\circ SK)(L(\mathcal{H}))\subset \mathcal{S}%
^{\prime }(\mathbb{R}^{2m})$ and $\gamma :(\mathcal{F}_{I}\circ \tau ^{\ast
}\circ SK)(L(\mathcal{H}))\rightarrow \mathcal{M\subset E}_{\mathbb{C}}(%
\mathbb{R}^{2m})$ defined by $\gamma (T)(q,p)=(\omega \ast T)(p,q)=\left[
(\omega \ast T)\circ \Xi \right] (q,p)$ . Note that $\mathcal{M}$ is
contained in the space of slowly increasing differentiable functions (cfr. 
\textbf{[V]} ch.I, par.5.6.c). Since $\left\langle A^{\ast }\right\rangle =%
\overline{\left\langle A\right\rangle }$, the image $\left\langle \cdot
\right\rangle (L_{sa}(\mathcal{H}))$ is the real part $\mathcal{M}_{\mathbb{R%
}\text{ }}$of $\mathcal{M}$.
\end{notation}

\begin{theorem}
For every bounded operator $A$ and every $(q,p)$ in $\mathbb{R}^{2m}$ we
have: $\left\langle A\right\rangle (q,p)=(2\pi )^{3m/2}\cdot \left\{ \omega
\ast \mathcal{F}_{I}\left[ SK(A)\circ \tau \right] \right\} (p,q)$ and if $%
SK(A)$ is regular 
\begin{equation*}
\left\langle A\right\rangle (q,p)=\frac{1}{\pi ^{m}}\cdot \int_{\mathbb{R}%
^{3m}}SK(A)(b+\frac{t}{2},b-\frac{t}{2})\cdot e^{-[\left\| p-a\right\|
^{2}+\left\| q-b\right\| ^{2}]}\cdot e^{-i\cdot t\cdot a}\cdot dt\cdot
da\cdot db\text{ \ \ }
\end{equation*}
\end{theorem}

\begin{proof}
$\left\langle \vartheta _{qp},A\vartheta _{qp}\right\rangle =(\pi
)^{-m}\cdot SK(A)_{rs}\left( \int_{\mathbb{R}^{m}}e^{-\left\| q-\frac{r+s}{2}%
\right\| ^{2}-\left\| a\right\| ^{2}}\cdot e^{-i\left( r-s\right) \left(
p-a\right) }\cdot da\right) $ and $(2\pi )^{3m/2}\cdot \left\{ \omega \ast 
\mathcal{F}_{I}\left[ SK(A)\circ \tau \right] \right\} (p,q)=$

$=(\pi )^{-m}\cdot SK(A)_{rs}\left( \int_{\mathbb{R}^{m}}e^{-\left\| q-\frac{%
r+s}{2}\right\| ^{2}-\left\| p-t\right\| ^{2}}\cdot e^{-i\left( r-s\right)
\cdot t}\cdot dt\right) $
\end{proof}

\begin{corollary}
The map $\left\langle \cdot \right\rangle =(2\pi )^{3m/2}\cdot \gamma \circ 
\mathcal{F}_{I}\circ \tau ^{\ast }\circ SK:L(\mathcal{H})\rightarrow 
\mathcal{M}$ is a $\mathbf{C-}$linear isomorphism with $\left\langle \cdot
\right\rangle ^{-1}=(2\pi )^{-3m/2}\cdot SK^{-1}\circ (\tau ^{-1})^{\ast
}\circ \mathcal{F}_{I}^{-1}\circ \gamma ^{-1}$
\end{corollary}

\begin{notation}
To avoid to deal with the ''exotic'' Fourier transform $\mathcal{F}[e^{\frac{%
1}{4}[\left\| x\right\| ^{2}+\left\| y\right\| ^{2}]}]$ we introduce a
cut-off. For each positive integer $N$ let's choose, once for all, a ''hat''
function $h_{N}$ in $\mathcal{D}(\mathbb{R}_{xy}^{2m})$ always between $0$
and $1$ with value $1$ on $_{{}}\prod_{1}^{2m}[-N,N]$ and value $0$ outside
of $\prod_{1}^{2m}[-N-1,N+1]$ and moreover pair and invariant under the
exchange of $x$ with $y$.
\end{notation}

\begin{theorem}
For each $G$ in $\mathcal{M}$ we have: 
\begin{equation*}
\gamma ^{-1}(G)(F)=(2\pi )^{m}\cdot \lim_{N\rightarrow \infty }G\left( 
\mathcal{F}[h_{N}\cdot e^{\frac{1}{4}[\left\| \cdot \right\| ^{2}+\left\|
\cdot \cdot \right\| ^{2}]}]\star (F\circ \Xi )\right)
\end{equation*}
on every $F$ in $\mathcal{S}(\mathbb{R}^{2m})$
\end{theorem}

\begin{proof}
Computation.
\end{proof}

\begin{definition}
For each $\varphi $ and $\psi $ in $\mathcal{H}$ let's define $S_{\varphi
\psi }:\mathcal{M\rightarrow }\mathbb{C}$ as $S_{\varphi \psi
}(G)=\left\langle \varphi ,[\left\langle \cdot \right\rangle ^{-1}(G)]\psi
\right\rangle $ (when $\varphi =\psi $ we will write $S_{\varphi }$ instead
of $S_{\varphi \varphi }$).
\end{definition}

\begin{remark}
Obviously $S_{\varphi \psi }$ is defined in such a way to have: 
\begin{equation*}
S_{\varphi \psi }(\left\langle A\right\rangle )=\left\langle \varphi ,A\psi
\right\rangle
\end{equation*}
\end{remark}

\begin{theorem}
If $\varphi $ and $\psi $ are in $\mathcal{S}(\mathbb{R}^{m})$ we have, for
every $G$ in $\mathcal{M}$ : 
\begin{equation*}
S_{\varphi \psi }(G)=(2\pi )^{-m/2}\lim_{N}\left[ \mathcal{F}\left(
h_{N}\cdot e^{\frac{1}{4}[\left\| \cdot \right\| ^{2}+\left\| \cdot \cdot
\right\| ^{2}]}\right) \star \left\{ \mathcal{F}_{I}^{{}}[(\psi \otimes 
\overline{\varphi })\circ \tau ]\circ \Xi \right\} \right] (G)
\end{equation*}
\end{theorem}

\begin{proof}
It is a straightforward application of the definition of $S_{\varphi \psi }$
and the formula for $\gamma ^{-1}(G)$
\end{proof}

\begin{exercise}
\begin{enumerate}
\item  For each $(q_{0},p_{0})$ in $\mathbb{R}^{2m}$ we have $S_{\theta
_{q_{0}p_{0}}}=\delta _{q_{0}p_{0}}$

\item  For each non-zero polynomial $P(x_{1}^{{}},...,x_{m})$ the map $%
S_{\theta \cdot P}$ is a distribution with support in $(0,0)$ (if $\psi $ is
in $\theta \cdot \mathbb{C}[x]$ then $\mathcal{F}_{II}^{-1}[(\psi \otimes 
\overline{\psi })\circ \tau ]\circ \Xi $ is in $e^{-\frac{1}{4}\left\| \cdot
\right\| ^{2}}\cdot \mathbb{C}[a,b]$).

\item  $S_{\sqrt{2}\cdot x_{1}\cdot \theta }(G)=\delta _{00}(G)+\frac{1}{2}%
(\partial _{u_{1}u_{1}}^{2}+\partial _{v_{1}v_{1}}^{2})\mid _{00}(G)$ and it
is not a non-negative distribution or a signed measure

\item  For every $\varphi $ and $\psi $ in $L^{2}(\mathbb{R}^{m})$ and $%
N\geq 1$ the function: 
\begin{equation*}
S_{\varphi \psi N}=(2\pi )^{-m/2}\mathcal{F}\left( h_{N}\cdot e^{\frac{1}{4}[%
\left\| \cdot \right\| ^{2}+\left\| \cdot \cdot \right\| ^{2}]}\right) \star
\left\{ \mathcal{F}_{I}^{{}}[(\psi \otimes \overline{\varphi })\circ \tau
]\circ \Xi \right\}
\end{equation*}

is a well defined differentiable function and a multiplier on $\mathbb{R}%
^{2m}$.
\end{enumerate}
\end{exercise}

\begin{theorem}
Given $\varphi $ and $\psi $ in $L^{2}(\mathbb{R}^{m})$ for every bounded
operator $A$ we have $\left\langle \varphi ,A\psi \right\rangle
=\lim_{N}S_{\varphi \psi N}(\left\langle A\right\rangle )=\lim_{N}\int_{%
\mathbb{R}^{2m}}S_{\varphi \psi N}\cdot \left\langle A\right\rangle \cdot
d\lambda $.
\end{theorem}

\begin{remark}
The terms $\left\langle A\right\rangle $ and $S_{\varphi \psi }$ are
connected with the Wigner and\ Husimi \ transforms: \ since the expression $%
(2\pi )^{-m/2}\mathcal{F}_{I}^{{}}[(\psi \otimes \overline{\varphi })\circ
\tau ]\circ \Xi $ corresponds to $\mathcal{W}^{1}(\psi ,\varphi )$ and $%
\mathcal{H}^{1}(\psi ,\varphi )=$ $2^{m}\cdot \mathcal{W}^{1}(\psi ,\varphi
)\star e^{-^{[\left\| \cdot \right\| ^{2}+\left\| \cdot \cdot \right\|
^{2}]}}$(cfr. \textbf{[K] }when $\varepsilon =1$ ) we have: $\mathcal{W}%
^{1}(\psi ,\varphi )=2^{m}\cdot e^{-^{[\left\| \cdot \right\| ^{2}+\left\|
\cdot \cdot \right\| ^{2}]}}\star S_{\varphi \psi }^{{}}$ and $\mathcal{H}%
^{1}(\psi ,\varphi )=e^{-\frac{1}{2}[\left\| \cdot \right\| ^{2}+\left\|
\cdot \cdot \right\| ^{2}]}\star S_{\varphi \psi }^{{}}$.

Note also the equality: $\left\langle E_{\psi }\right\rangle =(\pi
/2)^{m}\cdot \mathcal{H}^{1}(\psi ,\psi )$. Moreover it is not difficult to
prove that the Weyl operator $Op^{W}(a)$ associated to the symbol $a$ has: $%
\left\langle Op^{W}(a)\right\rangle =(2\pi )^{m}\cdot \omega \ast a$.
\end{remark}

\begin{notation}
Sometimes we will find useful to identify $(x,y)$ in \ $\mathbb{R}^{2m}$
with $z=x+iy$ in $\mathbb{C}^{m}$, $(q,p)$ with $w=q+ip$ etc. We will need
in the following the functions: $\Omega _{N}:\mathbb{C}^{m}\times \mathbb{C}%
^{m}\times \mathbb{C}^{m}\rightarrow \mathbb{C}$ and $\Delta :\mathbb{C}%
^{m}\times \mathbb{C}^{m}\times \mathbb{C}^{m}\rightarrow \mathbb{R}$ given
by: 
\begin{equation*}
\Omega _{N}(w,w^{\prime },w^{\prime \prime })=\frac{4}{(2\pi ^{2})^{m}}\cdot
e^{-\left\| w\right\| ^{2}}\cdot \mathcal{F}\mathit{(h}_{N}\cdot e^{\frac{1}{%
4}\left\| \cdot \right\| ^{2}}\mathit{)(w}^{\prime }\mathit{)}\cdot \mathcal{%
F}\mathit{(h}_{N}\cdot e^{\frac{1}{4}\left\| \cdot \right\| ^{2}}\mathit{)(w}%
^{\prime \prime }\mathit{)}
\end{equation*}

and 
\begin{equation*}
\Delta (w,w^{\prime },w^{\prime \prime })=\det \left[ 
\begin{array}{ccc}
1 & RE(w) & IM(w) \\ 
1 & RE(w^{\prime }) & IM(w^{\prime }) \\ 
1 & RE(w^{\prime \prime }) & IM(w^{\prime \prime })
\end{array}
\right]
\end{equation*}
\end{notation}

\begin{theorem}
\ For every couple $A$ and $B$ of bounded operators we have: 
\begin{equation*}
\left\langle A\cdot B\right\rangle (z)=\lim_{N}\int_{\mathbb{C}^{m}\times 
\mathbb{C}^{m}}\left\langle A\right\rangle (z^{\prime })\cdot \left\langle
B\right\rangle (z^{\prime \prime })\cdot (\Omega _{N}\ast e^{-2i\Delta
})(z,z^{\prime },z^{\prime \prime })\cdot dz^{\prime }\cdot dz^{\prime
\prime }
\end{equation*}
\end{theorem}

\begin{proof}
It is a lenghty but not difficult calculation of $\left\langle A\cdot
B\right\rangle (z)$.
\end{proof}

\ 

\begin{notation}
\ If we denote by $G\times H$ the function 
\begin{equation*}
(G\times H)(z)=\lim_{N}\int_{\mathbb{C}^{m}\times \mathbb{C}^{m}}G(z^{\prime
})\cdot H(z^{\prime \prime })\cdot (\Omega _{N}\ast e^{-2i\Delta
})(z,z^{\prime },z^{\prime \prime })\cdot dz^{\prime }\cdot dz^{\prime
\prime }
\end{equation*}
we have an associative product in $\mathcal{M}$ : such that: $\left\langle
A\cdot B\right\rangle =\left\langle A\right\rangle \times \left\langle
B\right\rangle $ \ (that is the map $\left\langle \cdot \right\rangle :(%
\mathcal{L}\mathit{(}\mathcal{H}\mathit{),+,\cdot })\mathit{\rightarrow }(%
\mathcal{M}\mathit{,+,\times })$ is an isomorphism of algebras).

We will denote by $\left\{ H,G\right\} $ the expression $i\cdot \left(
H\times G-G\times H\right) $ given by the function 
\begin{equation*}
(\left\{ H,G\right\} )(z)=2\cdot \lim_{N}\int_{\mathbb{C}^{m}\times \mathbb{C%
}^{m}}H(z^{\prime })\cdot G(z^{\prime \prime })\cdot \left[ \Omega _{N}\ast
\sin (2\cdot \Delta )\right] (z,z^{\prime },z^{\prime \prime })\cdot
dz^{\prime }\cdot dz^{\prime \prime }
\end{equation*}
\end{notation}

\section{Bibliography}

\ \ \ \ \ 

\textbf{[CZ]\ }\ T.L. Curtright and C.K. Zachos: Quantum mechanics in phase
space.World Scientific - Singapore (2005)

\textbf{[dG]} \ \ M.A. de Gosson: Symplectic geometry, Wigner-Weyl-Moyal
calculus, and quantum mechanics in phase space. - Pre. Univ. Potsdam - 2006

\textbf{[Gr]} \ \ H.J. Groenewold: On the principles of elementary quantum
mechanics. Physica 12 (1946) 405-460

\textbf{[K]} \ \ \ \ J.F. Keller: Computing semiclassical quantum
expectations by Husimi functions. Master Thesis - Technische Universitat
Munchen - 2012

\textbf{[M] }\ \ \ J.E.Moyal: Quantum mechanics as a statistical theory. \
Proc. Cambridge Phil. Soc. 45 (1949) 99-124

\textbf{[P]} \ \ \ \ J. C. T. Pool: Mathematical aspects of the Weyl
correspondence. Jour. of Math. Phys. Vol. 7 N. 1 Jan. 1966

\textbf{[R]} \ \ \ \ H. L. Royden: Real analysis - The Macmillan company -
London 1968

\textbf{[T]} \ \ \ \ F. Treves: Topological Vector Spaces, Distributions and
Kernels. - Academic Press, NY 1967

\textbf{[V]} \ \ \ V. S. Vladimirov: Generalized functions in Mathematical
physics - USSR 1979

\textbf{[We]} \ \ J. Weidmann: Linear Operators in Hilbert Spaces.
Springer-Verlag, NY 1980

\end{document}